\documentclass[sigconf]{acmart}

\usepackage{float}
\usepackage{xurl}
\hypersetup{hidelinks}
\usepackage{listings}
\usepackage[dvipsnames]{xcolor}
\usepackage{graphicx}
\usepackage{booktabs}
\usepackage{array}
\usepackage{mdframed}
\usepackage[most]{tcolorbox}
\usepackage{soul}
\usepackage{url}
\usepackage{hyperref}
\usepackage{environ}
\usepackage{colortbl}
\usepackage{multirow}
\usepackage{natbib}
\usepackage{duckuments}

\definecolor{myyellow}{HTML}{FFC001}
\definecolor{myblue}{HTML}{2878B5}

\usepackage{tikz}

\newcommand*\circnum[2][myblue]{%
  \tikz[baseline=(n.base)]%
  \node[
    draw=white,
    fill=#1,
    text=white,
    circle,
    minimum size=9pt,
    inner sep=0pt,
    font=\sffamily\bfseries\scriptsize
  ] (n) {#2};%
}

\renewcommand\footnotetextcopyrightpermission[1]{} 

\setcopyright{none}

\acmConference[VEM 2026]{14th Workshop on Software Visualization, Maintenance and Evolution}{September 11, 2026}{São Paulo, SP, Brazil}

\AtBeginDocument{
    
}

\begin{document}

\title{PrivDev: Mapping Static-Analysis Data Types to DPV}

\author{Simon Bernbeck}
\affiliation{
  \institution{PUC-Rio}
  \city{Rio de Janeiro}
  \country{Brazil}
}
\email{sbernbeck@inf.puc-rio.br}

\author{Ricardo Ramalho}
\affiliation{
  \institution{PUC-Rio}
  \city{Rio de Janeiro}
  \country{Brazil}
}
\email{rjunior@inf.puc-rio.br}

\author{Matheus Amendoeira}
\affiliation{
  \institution{PUC-Rio}
  \city{Rio de Janeiro}
  \country{Brazil}
}
\email{mamendoeira@inf.puc-rio.br}

\author{Juliana Alves Pereira}
\affiliation{%
  \institution{PUC-Rio}
  \city{Rio de Janeiro}
  \country{Brazil}
}
\email{juliana@inf.puc-rio.br}

\renewcommand{\shortauthors}{Bernbeck et al.}
\renewcommand{\shorttitle}{PrivDev: Mapping Data Types to DPV}

\begin{abstract}

Static-analysis scanners can identify personal-data types in source code, but they lack mechanisms to connect these findings to standardized privacy vocabularies. PrivDev maps 122 Bearer CLI data types to Data Privacy Vocabulary Personal Data (DPV-PD) categories and links them to potentially relevant GDPR provisions. Our approach combines deterministic mapping for 43 exact-label matches with a retrieval-grounded Large Language Model (LLM) to resolve the remaining 79 non-trivial mappings.
The resulting RDF knowledge graph contains 118 ODRL policy resources that were structurally validated using SHACL. The artifact passed five complementary validation gates that cover structural correctness, query consistency, retrieval quality, LLM-based assessment, and human evaluation. In human evaluation, nine annotators produced 711 judgments, yielding a raw agreement of 0.72, Gwet's AC1 of 0.68, and Gwet's AC2 of 0.88. Our results indicate that the proposed mappings are plausible and reproducible, while also revealing ambiguities in scanner-defined data-type labels and coverage gaps in DPV-PD.

\end{abstract}


\frenchspacing
\maketitle

\section{Introduction}\label{sec:intro}
Software developers face a growing burden to embed privacy compliance into the systems they build. Regulations such as the European Union's General Data Protection Regulation (GDPR) impose concrete obligations on controllers and processors, yet the translation from legal text to implementation practice remains largely unsupported by tooling. Recent empirical work \citep{matos2025} surveyed 88 Brazilian developers and found that most conflate privacy with security, and that 89.74\% of teams handling privacy ad-hoc report using no privacy tools. This finding is consistent with earlier findings showing that developers struggle to embed privacy into software without dedicated tooling and mindset support \cite{senarath2018,hadar2018}. They reveal a structural gap: \textit{developers write the code but do not know the law, while privacy lawyers understand GDPR but cannot read code}.

Security scanners such as Bearer CLI \cite{bearer} use static analysis to detect data types in source code, \textit{e.g.}, identifying that an email address flows through a code path. However, they do not connect those data types to a privacy vocabulary or explain which obligations may warrant review. In contrast, the Data Privacy Vocabulary (DPV) \cite{pandit2025,w3cdpv} provides a machine-readable taxonomy of personal data categories, processing purposes, and legal bases, but has no direct connection to static-analysis outputs. Our literature search \citep{pandit2025, Ran2025GDPR-Bench-Android:,Ferreira2023RuleKeeper:} found no approach that directly maps static-analysis datatype labels to DPV categories and links them to potentially relevant GDPR provisions. 

This paper presents PrivDev, a security-to-privacy bridge that maps 122 Bearer CLI datatype taxonomy entries to DPV personal data (DPV-PD) categories (\texttt{pd:*}) and connects these categories to potentially relevant GDPR provisions. Methodologically, we integrate Design Science Research (DSR) \cite{hevner2004}, the NeOn methodology \cite{suarez2012}, and the Framework for Evaluation in Design Science Research (FEDS) \cite{venable2016} to guide the design, ontology re-engineering, validation, and evaluation of PrivDev. This integration provides a systematic approach for transforming an existing static-analysis taxonomy into a machine-readable privacy artifact and evaluating it through complementary automated and human evidence.

The main contributions of this paper are:
    (1) We propose a two-stage mapping strategy that combines deterministic matching for 43 exact-label correspondences with retrieval-grounded Large Language Model (LLM) use for 79 non-trivial mappings, enabling the systematic treatment of heterogeneous scanner-defined datatype labels.
    (2) We provide an RDF knowledge graph containing 118 ODRL policy resources and structurally validated using SHACL. The resulting artifact supports SPARQL-based queries that connect static-analysis data types with DPV categories and potentially relevant GDPR provisions.
    (3) We evaluate the artifact through five complementary validation gates covering structural correctness, SPARQL query consistency, retrieval quality, LLM-based assessment, and human evaluation. The human evaluation comprises 711 judgments from nine annotators and provides evidence of mapping plausibility while revealing ambiguities in scanner-defined datatype labels and coverage gaps in DPV-PD.

\textit{Target audience.} Our study supports software developers and privacy practitioners seeking to connect source-code analysis with privacy requirements.

\vspace{-4pt}
\section{Background and Related Work}\label{sec:review}
The Data Privacy Vocabulary (DPV) \cite{pandit2025,w3cdpv}, maintained by the W3C Data Privacy Vocabularies and Controls Community Group, provides a machine-readable taxonomy of privacy concepts. It covers personal data categories (\texttt{pd:*}), processing purposes (\texttt{pu:*}), legal bases (\texttt{lb:*}), technical and organisational measures, rights, and other concepts. Its personal-data taxonomy (\texttt{pd:*}) comprises 235 classes in DPV 2.3, which are organised under top-level categories (\textit{e.g.}, financial, health, contact, identification, location). DPV is formalised in OWL~2 (Web Ontology Language 2)\footnote{OWL~2 is a W3C standard for representing ontologies and their semantic relationships in a machine-readable form.} and extensible through DPV-PD, DPV-OBL, and other modules. The DPV-OBL \cite{dpvcg2025rules} extension used in this work associates GDPR articles with data categories and processing purposes. We inherit these associations as review cues, not legal determinations of applicability, without independently validating them against legal doctrine or case law.

This work encoded the derived privacy rules using three complementary W3C standards: ODRL, SHACL and SPARQL. The Open Digital Rights Language (ODRL)~\cite{odrl} expresses permissions, prohibitions, and duties over digital assets (\textit{e.g.}, if a variable of type \texttt{pd:Financial} is processed, the duty of conducting a Data Protection Impact Assessment (DPIA) must be fulfilled). Each Bearer data type produces at least one ODRL rule under our approach. SHACL (Shapes Constraint Language)~\cite{shacl} validates the resulting policies against node shapes that restrict which ODRL properties are allowed for each DPV category (\textit{e.g.}, a \texttt{pd:Health} mapping may expose \texttt{lb:ExplicitConsent}; a \texttt{pd:Financial} mapping may expose retention and DPIA duties). SPARQL \cite{w3csparql} is used to write competency questions, expressed as ASK or SELECT queries, that verify each generated policy against its expected response.

Pandit and Esteves \cite{pandit2024duo} present the closest precursor to our work. They manually transform DUO (Data Use Ontology) permission and modifier codes into machine-readable ODRL~+~DPV policies, representing data-use restrictions, dataset policies, requests and decisions as ODRL instances and expressing privacy and legal-compliance concepts using DPV. Their validation is a proof-of-concept demonstration rather than formal schema validation. Moreover, applying a similar manual alignment process to each static-analysis scanner or personal-data detection tool would require a separate alignment effort \cite{Tokas2021Static, Baramashetru2025A,Khedkar2023Static,Khedkar2024Toward}, limiting scalability across heterogeneous data-type taxonomies. This motivates our retrieval-grounded, LLM-assisted approach, which supports adaptation to other data-type taxonomies\footnote{The present evaluation remains limited to Bearer's 122 data types.}. Khedkar et al. \cite{khedkar_between_2026} extend this human-centered perspective by mapping Android source-code slices to DPV concepts to support privacy assessors during DPIAs. Their approach relies on manual curation and was evaluated through interviews with 16 participants. Instead, our pipeline automatically generates and structurally verifies ODRL/SHACL artifacts, reducing the manual effort required to establish and validate these mappings.

\vspace{-3pt}
\section{Methodology}\label{sec:methology}
This study focuses on \emph{PrivDev Layer~1}, the first layer of the proposed security-to-privacy bridge. Layer~1 maps personal-data types detected by the Bearer CLI static-analysis tool to the corresponding categories of personal data in the DPV and connects these categories to potentially relevant GDPR provisions. Future layers may extend this foundation to security weaknesses and broader privacy risks, connecting Common Weakness Enumeration (CWE) security findings to the privacy layer and further relating these findings to risks defined in the OWASP Security and Privacy Top~10.

The methodology combines three frameworks: DSR, NeOn, and FEDS. Figure~\ref{fig:layer1-pipeline} shows how the pipeline instantiates these frameworks, and the following sections describe their roles and implementation in detail.

\vspace{-3pt}
\subsection{Research Design: DSR, NeOn, and FEDS}

This work adopts Hevner's Design Science Research (DSR) framework~\cite{hevner2004}, which treats a research contribution as an \emph{artifact} that must be rigorously evaluated. The Layer~1 bridge repurposes an existing resource for a new context, an \emph{exaptation} in DSR terms: the Bearer taxonomy was built for security scanning rather than privacy reasoning, and re-encoding it in ODRL~+~DPV adapts it to this new purpose. The artifact is both a model, defining a Bearer-to-DPV mapping, and an instantiation, an executable pipeline that generates, validates, and queries that mapping.

NeOn~\cite{suarez2012} provides the construction method, 
\emph{Reusing and Re-engineering Non-Ontological Resources}, which covers transforming classification schemes and thesauri into ontologies: here, the 122-entry Bearer taxonomy into an ODRL~+~DPV Turtle artifact linking each Bearer type to a DPV \texttt{pd:*} category.

FEDS~\cite{venable2016} structures the evaluation. Neither Bearer nor DPV publishes an authoritative Bearer-to-\texttt{pd:*} alignment, so no ground truth exists. We therefore adopt a convergent-evidence evaluation design that combines automated structural checks (SHACL, OOPS!, and SPARQL competency questions), an LLM-judge quality assessment (DeepEval), and independent human judgment on the non-trivial mappings. Evaluation is performed during artifact construction and after the artifact is complete.

\vspace{-3pt}
\subsection{Construction Procedure}
\label{sec:construction-procedure}

\begin{figure}[!h]
    \centering
    \includegraphics[width=.44\textwidth]{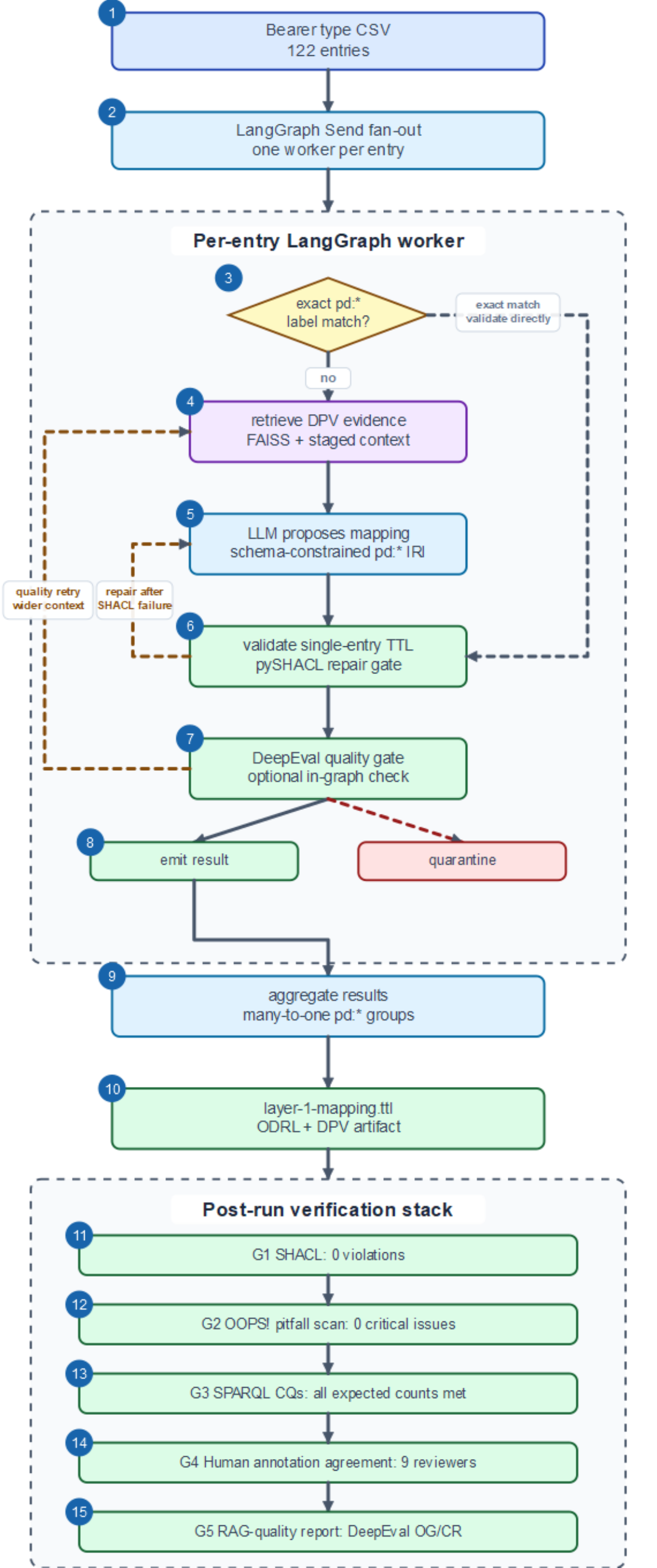}
    \vspace{-5pt}
    \caption{Layer~1 construction pipeline.}
    \Description{Vertical flow chart from Bearer CSV input through LangGraph mapping, validation, aggregation, and gates G1--G5, with each box numbered 1 through 15 in execution order.}
    \label{fig:layer1-pipeline}
    \vspace{-15pt}
\end{figure}

Figure~\ref{fig:layer1-pipeline} shows the construction procedure's execution order, with each box numbered (1--15); the text below follows the same flow and cites the step numbers.

\vspace{-3pt}
\subsubsection{Input Taxonomy and Per-Entry Dispatch}

The procedure starts from a Bearer data-type CSV containing 122 personal-data types~\circnum{1}. A LangGraph outer graph dispatches the entries in parallel~\circnum{2}, with one worker resolving each Bearer type. This preserves an auditable trace per data type while independent mappings run concurrently.

Each worker first checks whether the Bearer type label matches exactly a \texttt{pd:*} \texttt{skos:prefLabel} \circnum{3}. These lexically unambiguous cases are resolved deterministically from the DPV taxonomy and skip retrieval and LLM proposal. They are still validated before emission, but are reported separately as \texttt{trivial\_pass} entries because they do not require semantic classification.

\vspace{-3pt}
\subsubsection{Evidence-Based Mapping}

Prior work on tool-augmented ontology alignment finds that LLMs provided with raw OWL files can perform worse than to give the LLM access to the ontology through a retrieval/tool mechanism \cite{rovai2026openontologies}. We therefore retrieve structured, per-concept chunks containing the label, \texttt{skos:broader} hierarchy, definition, and scope note, rather than passing the raw DPV Turtle to the mapping LLM \circnum{4}. Retrieval uses three FAISS~\cite{faiss} indexes. \texttt{ontology\_target} contains DPV-PD concepts and provides the candidate classes from which the LLM selects. \texttt{detector\_aliases} contains synonym labels from personal-data detection tools, while \texttt{sector\_schema} contains domain-specific field aliases; both provide supporting evidence when a Bearer label lacks a close DPV match. A two-stage retrieve-and-rerank process first uses FAISS to retrieve candidates, and then applies a cross-encoder to rank the candidates and retain a small set for the LLM. The replication package lists the index sizes, embedding model, and top-$k$ values.

\vspace{-3pt}
\subsubsection{Structured Mapping Proposal}

For non-trivial entries, the LLM \texttt{deepseek-v4-flash} proposes the mapping \circnum{5}. We selected this model over GPT-4o-mini because of its higher rate limit and lower per-token cost; the task did not require flagship-model reasoning. The prompt provides a compact list of \texttt{pd:*} IRI (Internationalized Resource Identifier), the retrieved evidence grouped into four sections (primary DPV candidate, alternatives, synonym aliases, sector-schema confirmations), and the instruction to prefer an exact label correspondence over a more specific or broader class.

The LLM does not generate Turtle. Instead, it returns a JSON object with three fields: \texttt{pd\_iri}, \texttt{confidence} (high/medium/low), and \texttt{notes} (the verbatim DPV definition supporting the selection). The response is validated against a predefined schema before the artifact is generated. GDPR article IRIs are derived deterministically from the selected \texttt{pd:*} class rather than generated by the LLM.

\subsubsection{Policy Encoding, Validation, and Aggregation}

The mapping is encoded as an ODRL policy through deterministic templates. For each (Bearer type, DPV class) pair, an \texttt{odrl:Policy} of type \texttt{odrl:Set} targets the Bearer data type to the corresponding DPV obligation and ODRL constraints from the DPV-OBL extension.

Each generated policy is validated in-graph \circnum{6} against the SHACL shapes detailed as Gate G1 in Section~\ref{sec:formative-gates}. After a violation, its text is appended to the LLM prompt and the worker retries, with up to 3 attempts before quarantine. A policy that passes SHACL proceeds to the in-graph DeepEval quality check \circnum{7}, detailed as Gate G5 in Section~\ref{sec:formative-gates}. An ontology-grounding failure triggers blocking after retry, widening, and reformulating retrieval before another proposal. Entries that clear both checks are emitted \circnum{8}; those that exhaust the repair or quality budget are quarantined.

The outer graph aggregates the results per-entry in the final Turtle mapping artifact \circnum{9}. Entries with the same selected \texttt{pd:*} IRI are grouped in a many-to-one policy. This produces the final ODRL~+~DPV artifact used by the query layer \circnum{10}. Five post-run gates then evaluate the consolidated artifact: G1 (SHACL \circnum{11}), G2~(OOPS!~\circnum{12}), G3 (SPARQL competency questions \circnum{13}), G4 (Human annotation \circnum{14}), and G5 (DeepEval \circnum{15}), as detailed in Section~\ref{sec:all-gates}.

\subsection{FEDS Evaluation Design} \label{sec:all-gates}

\subsubsection{Formative Verification Gates}
\label{sec:formative-gates}

Table \ref{tab:formative-gates} describes the five formative gates. They provide complementary evidence rather than a single measure of correctness. Gates G1--G3 assess the structural integrity and query consistency of the artifact, while G4 and G5 provide evidence about the semantic adequacy of the non-trivial mappings through independent human assessment and retrieval-grounded evaluation.

\begin{table}[H]
\centering
\vspace{-5pt}
\caption{Formative verification gates for PrivDev Layer~1.}
\vspace{-5pt}
\label{tab:formative-gates}
\footnotesize
\begin{tabular}{@{}p{1.3cm} p{6.65cm}@{}}
\toprule
\textbf{Gate/Method} & \textbf{Purpose, scope, and pass criterion}\tabularnewline
\midrule

\textbf{G1} SHACL &
Validates the structural well-formedness of generated ODRL policies against three shapes: class well-formedness, required policy links, and Article~9 consistency. Validation is performed during generation and exhaustively on the consolidated artifact. Pass: zero violations.\tabularnewline
\addlinespace[1.25pt]

\textbf{G2} OOPS! &
Scans the consolidated RDF artifact against the complete default OOPS! pitfall catalog~\cite{poveda2014}. Findings are classified using OOPS!'s Critical, Important, and Minor severity levels. Pass: zero Critical findings; Important and Minor findings are recorded but non-blocking.\tabularnewline
\addlinespace[1.25pt]

\textbf{G3} SPARQL &
Checks structural consistency of the consolidated artifact through five competency questions (CQ-01--05$^*$), covering mapping coverage, Article~9 links, policy links, many-to-one grouping, and DPV namespace validity. It does not assess whether an individual Bearer type is mapped to the semantically correct DPV class. Pass: each CQ returns its expected row count (122 for CQ-01, 0 for CQ-02--05).\tabularnewline
\addlinespace[1.25pt]

\textbf{G4} Human &
Assesses the 79 non-trivial Bearer-to-DPV mappings through independent human judgments (Section~\ref{sec:annotation-method}). Annotators classify mappings as \textit{Correct}, \textit{Too broad}, \textit{Too narrow}, \textit{Wrong}, or \textit{Unclear}. Pass: raw agreement and Gwet's AC2 both $\geq 0.70$ (see Section \ref{sec:annotation-method}).\tabularnewline
\addlinespace[1.25pt]

\textbf{G5} DeepEval &
Evaluates the retrieval-grounded mapping process for non-trivial entries using Ontology Grounding (OG) and Contextual Relevancy (CR). OG checks whether the selected \texttt{pd:*} concept is supported by the retrieved evidence; CR assesses the relevance of the retrieved context. Pass: OG is the blocking criterion and can trigger retrieval reformulation or quarantine, while CR is diagnostic.\tabularnewline

\bottomrule
\end{tabular}
\scriptsize{$^*$The complete list of competency questions (CQ-01--05) is in our supplementary material.}
\vspace{-10pt}
\end{table}

\subsubsection{Gate G4: Human Annotation Protocol}
\label{sec:annotation-method}

A mixed panel of software developers and law practitioners reviewed the 79 non-trivial mappings using a KoboToolbox form\footnote{https://www.kobotoolbox.org/}. The remaining 43 exact-label matches were resolved deterministically and excluded from human annotation; this does not assume that lexical identity guarantees semantic correctness. For each entry, an annotator selects one of five ordinal judgments: \textit{Correct}, \textit{Too broad}, \textit{Too narrow}, \textit{Wrong}, or \textit{Unclear}. Annotators also rate confidence from 1 to 5. Confidence only flags the rows for qualitative review (Section~\ref{sec:results-annotation}). All valid annotators are pooled in one quantitative analysis. The self-reported personal-data-classification experience is also retained as descriptive metadata (Section~\ref{sec:results-annotation}) but does not determine inclusion.

\paragraph{Calibration, disagreement handling, and panel.} Before annotating, all respondents received a short recorded walkthrough of the research context, DPV/GDPR/LGPD background, and the form, plus written instructions covering judgment-scale definitions, worked examples (e.g. \textit{Email Address} $\to$ \texttt{pd:Identifier} as \textit{too broad}), and a quality-flag legend; no separate pilot-rating round was run. Annotators judged independently and once each, with no consensus or adjudication round; disagreement is characterized rather than resolved, statistically through the four reliability measures below and qualitatively through coded review of comments and \textit{Unclear} verdicts (Section~\ref{sec:results-annotation}). The panel combined software-engineering practitioners and students with law students and legal practitioners; some had prior personal-data-classification experience.

\textit{Too broad} and \textit{too narrow} are not equidistant from \textit{Correct}. An over-broad DPV class risks mis-claiming or omitting a GDPR Article~9 special-category obligation, which we treat as more severe than an over-narrow class. This is a task-specific modeling choice, not a universal legal rule. An ordinal severity scale $s$ maps the four verdicts to $[0,1]$:
\begin{equation}
\begin{aligned}
s(\text{Correct}) &= 1, & s(\text{Too narrow}) &= \tfrac{2}{3}, \\
s(\text{Too broad}) &= \tfrac{1}{3}, & s(\text{Wrong}) &= 0,
\end{aligned}
\label{eq:severity}
\end{equation}
\textit{Unclear} judgments are excluded from all quantitative statistics but counted and routed to qualitative review (Section~\ref{sec:results-annotation}). Such a verdict can indicate rater uncertainty or a genuine DPV coverage gap; a numeric placeholder would misrepresent both.

With $q=4$ scored categories and a correct-heavy marginal distribution, chance-corrected statistics such as Krippendorff's $\alpha$ can read paradoxically low despite high raw agreement~\cite{feinstein1990,gwet2008} (quantified in Section~\ref{sec:results-annotation}); we therefore report four complementary statistics. Let $n_{ik}$ be the number of raters assigning item $i$ to category $k \in \{1,\ldots,q\}$, and let $r_i$ be the number of non-\textit{Unclear} judgments on item $i$. Items with $r_i < 2$ do not contribute to a pairwise statistic; $N$ is the number of items with $r_i \geq 2$.

\paragraph{Raw percent agreement} is the mean pairwise exact-match rate over all unordered rater pairs $(a,b)$ across qualifying items:
\begin{equation}
P_a^{\text{raw}} = \frac{1}{\sum_i \binom{r_i}{2}} \sum_{i}\ \sum_{\{a,b\} \subseteq \text{raters}(i)} \mathbb{1}\!\left[v_{ia} = v_{ib}\right],
\label{eq:pa-raw}
\end{equation}
where $v_{ia}$ is rater $a$'s verdict on item $i$.

\paragraph{Krippendorff's $\alpha$ (ordinal) and Gwet's AC1 (nominal, prevalence-robust)} are two chance-corrected statistics computed on the same judgments. $\alpha$~\cite{krippendorff2004} corrects for chance using the observed marginal distribution, and is known to read low under prevalence skew even when raw agreement is high. Gwet's AC1~\cite{gwet2008,gwet2014} corrects for chance using the observed \emph{agreement} propensity instead, making it far less sensitive to that same skew. Both reduce to a $(P_a - P_e)/(1 - P_e)$ form over the per-item category counts $n_{ik}$; full derivations are in Krippendorff~\cite{krippendorff2004} and Gwet~\cite{gwet2008,gwet2014}.

\paragraph{Gwet's AC2 (weighted, ordinal + prevalence-robust)} extends AC1 with the same quadratic distance weights used for $\alpha$, combining prevalence robustness with severity-aware scoring:
\begin{equation}
w_{kl} = 1 - \left(\frac{s_k - s_l}{s_{\max} - s_{\min}}\right)^2,
\label{eq:gwet-w}
\end{equation}
applied to AC1's $P_a$ and $P_e$ terms above in place of the unweighted category counts and category propensities --- AC2, not $\alpha$, is read alongside raw agreement and AC1 when judging Gate G4's reliability pass criterion (Section~\ref{sec:results-annotation}).

\subsubsection{Internal Summative Checks}
\label{sec:summative-gates}

Once the artifact is complete, three internal summative checks close the evaluation: contribution completeness against an author-defined rubric, method reproducibility through a clean-environment rerun, and a self-assessment rubric covering completeness, correctness, consistency, clarity, and reproducibility. These checks assess process completeness, not independent effectiveness (Section~\ref{sec:gate-summary}).

\subsection{Research Questions}
    
This study is guided by three research questions (RQs), as follows:

\textbf{RQ$_1$.} \textit{Can the NeOn Classification-Scheme-to-OWL pattern, informed by the DUO/ODRL/DPV encoding mechanics of Pandit and Esteves \cite{pandit2024duo}, be adapted to produce a structurally valid mapping from Bearer CLI data-type detections to DPV \texttt{pd:*} personal-data categories?}
This question answers whether a pattern applied manually transfers to larger-scale automated data-type mapping.

\textbf{RQ$_2$.} \textit{Can the resulting SPARQL-queryable, DPV-grounded knowledge graph reliably retrieve potentially relevant GDPR provisions from Bearer CLI findings?}
This question answers whether the mappings carry enough structural and semantic grounding to reliably link findings to relevant provisions.

\textbf{RQ$_3$.} \textit{Can a FEDS-structured, convergent-evidence protocol provide a reproducible and sufficiently sensitive basis for evaluating the mappings in the absence of an established ground truth?}
This question answers whether converging structural, LLM-judge, and human evidence can credibly assess mapping quality.



\section{Implementation}\label{sec:impl}
\paragraph{Knowledge Graph Construction.}
The Layer~1 artifact is constructed by a LangGraph-orchestrated
pipeline that reads the 122-entry Bearer taxonomy, produces ODRL
policies backed by DPV categories, and emits a consolidated,
SHACL-validated knowledge graph together with a per-entry audit log.

\textit{Runtime components.} The two-level LangGraph structure
(outer dispatch, inner per-entry mapping/validation/quality
routing) is described in Section~\ref{sec:construction-procedure}. Each run writes
the consolidated Turtle artifact, the per-entry pipeline log, and the
RAG-quality report used by the evaluation section.

\textit{Artifact outputs.} The primary output is
\texttt{layer-1-mapping.ttl}, a query-visible ODRL graph backed by DPV. The pipeline also writes a JSON audit log for each processed Bearer type. It retains the parsed mapping decision (\texttt{pd:*} IRI, confidence, quality class, and GDPR article IRIs), the LLM's verbatim-quoted rationale from the structured response, the retrieved context chunks passed to the mapping LLM, and the DeepEval scores and verdicts. The raw unparsed LLM completion is not stored; post-hoc reproducibility checks use the parsed fields and retrieved context.

\textit{Cost.} The full 122-entry run completes in approximately 63
minutes wall-clock against the mapping LLM's hosted API, at approximately
\$0.37 in total token cost across roughly 1.91M tokens.

\paragraph{End-to-End Demonstration.}
To demonstrate the complete pipeline from the source code to potentially
relevant GDPR provisions, we
provide an end-to-end demonstration using OWASP Juice
Shop\footnote{\url{https://owasp.org/www-project-juice-shop/}} as the
target application (artifact and replication instructions:
\hyperref[sec:artifacts]{Artifact Availability}).
A Bearer CLI dataflow scan over the Juice Shop source tree maps each personal data type detection to its source location: file, line number, field name, and enclosing object. This provenance is additive because the bridge maps only data-type names; source locations decorate the output without affecting obligation retrieval. The consolidated graph is loaded into an RDF triple store with DPV 2.3. A SPARQL SELECT query resolves each Bearer label to its \texttt{pd:*} category and inherited DPV-OBL provision links in one round trip. A REST API exposes this as a repository scan using a public URL. The target codebase is cloned into an isolated temporary environment, scanned, matched against the graph, and discarded before results are returned, so no user source
code is retained. The query interface returns potentially relevant
GDPR-provision badges and collapsible source-location provenance for a Juice Shop scan.

\section{Results and Discussion}\label{sec:results}
\subsection{Mapping Coverage and Output}
\label{sec:mapping-coverage}

The Layer~1 pipeline processed all 122 Bearer data-type taxonomy
entries. All 122 (100\%) were mapped to DPV \texttt{pd:*} categories;
none were quarantined. These mapped types collapse to 118 distinct
\texttt{pd:*} targets (4 pairs share a category via many-to-one
grouping). The pipeline emits one consolidated Turtle graph containing
118 \texttt{odrl:Set} policy resources, with Bearer datatypes that share
a DPV-PD target grouped into the same policy. Table~\ref{tab:coverage}
summarises coverage, inherited DPV-OBL-derived provision links, and mapping-quality class. Applicability depends on the processing operation and purpose, not on the data category alone.

\begin{table}[H]
\centering
\small
\vspace{-5pt}
\caption{Layer~1 coverage: GDPR links and quality classes.}
\vspace{-5pt}
\label{tab:coverage}
\begin{tabular}{@{}lrr@{}}
\toprule
\multicolumn{3}{l}{\textit{DPV-OBL-derived GDPR provision links}} \\
\texttt{eu-gdpr:A6-1-a} (standard) & 104 & 85\% \\
\texttt{eu-gdpr:A9-2-a} (special category) & 18 & 15\% \\
\texttt{eu-gdpr:A10} (criminal-data-related) & 3 & 2\% \\
\multicolumn{3}{@{}p{0.95\columnwidth}@{}}{\footnotesize Rows are non-exclusive; a datatype may be linked to multiple provisions.} \\
\midrule
\multicolumn{3}{l}{\textit{Mapping quality class (of 122 mapped types)}} \\
\texttt{trivial\_pass} (exact label match) & 43 & 35\% \\
\texttt{auto\_pass} (accepted by automated checks) & 69 & 57\% \\
\texttt{flag\_revision} (emitted; marked for review) & 10 & 8\% \\
\bottomrule
\end{tabular}
\vspace{-10pt}
\end{table}


Across the 158 DeepEval judge calls issued for the 79 non-trivial entries
(two metrics per entry), mean ontology grounding was 0.941 and mean
contextual relevancy was 0.780.
Gate results (SHACL, OOPS!, SPARQL competency queries, RAG-quality, and
human-annotation agreement) are summarised in
Section~\ref{sec:gate-summary}.

\subsection{Gate G4: Human Annotation Results}
\label{sec:results-annotation}

Nine respondents reviewed all 79 non-trivial entries via the protocol
described in Section~\ref{sec:annotation-method} and are pooled into
one quantitative analysis (Table~\ref{tab:g4-reliability}). 

\begin{table}[H]
\small
\centering
\vspace{-10pt}
\caption{Gate G4 annotation reliability.}
\vspace{-10pt}
\label{tab:g4-reliability}
\begin{tabular}{lcc}
\toprule
Metric & All entries & Art.~9 ($n{=}10$) \\
\midrule
Raw percent agreement ($P_a^{\text{raw}}$) & 0.721 & 0.800 \\
Krippendorff's $\alpha$ (ordinal)          & 0.251 & 0.056 \\
Gwet's AC1 (nominal)                        & 0.682 & 0.785 \\
Gwet's AC2 (weighted)                       & 0.877 & 0.877 \\
\bottomrule
\end{tabular}
\vspace{-10pt}
\end{table}

The four statistics defined in Section~\ref{sec:annotation-method}
(Equations~\ref{eq:pa-raw}--\ref{eq:gwet-w}) 
disagree sharply on the same data:
72.1\% raw agreement versus an $\alpha$ of 0.251 --- the
high-agreement/low-kappa paradox~\cite{feinstein1990,gwet2008}, not a
contradiction. Judgments are heavily skewed toward \textit{Correct}
(558/706 non-Unclear judgments), and chance corrections based on the
observed marginal distribution fall sharply under this prevalence skew.
Gwet's AC1/AC2 are designed to resist this distortion. Their values of
0.68--0.88 are consistent with raw agreement rather than $\alpha$.
Scoring \textit{Unclear} as a disagreement category instead of excluding
it barely changes raw agreement (0.711 overall and 0.800 for the Article~9
subset), so the exclusion does not inflate the reported figure. Gate G4's
criterion, fixed before annotation, requires raw agreement and Gwet's AC2
to both be $\geq 0.70$; both thresholds are met (0.721 and 0.877 overall).
Krippendorff's $\alpha$ is not part of the criterion and did not determine
the gate outcome; we report it for transparency. Its low Article~9 value
($\alpha = 0.056$, $n{=}10$) reflects genuine disagreement in a small,
high-stakes sample, consistent with the prevalence-skew instability noted
above rather than a data artifact. We flag it for downstream attention
rather than treating it as resolved, and report Gate G4 as passing its
defined criterion.

\paragraph{What this validates.} Annotator agreement speaks to the
\emph{plausibility} of each Bearer-datatype-to-DPV-PD mapping, not to
full GDPR compliance: it does not confirm legal-basis correctness,
DPIA applicability, purpose-specific Article 9 escalation, or anything
about the Layers~2--3 mappings this paper does not cover.

\paragraph{Disagreement patterns.}

Comments, \textit{Unclear} verdicts, and low-confidence judgments
(78 of 711, extracted programmatically) reveal three recurring patterns,
unscored in Table~\ref{tab:g4-reliability}, motivating the DPV
coverage-gap discussion in Section~\ref{sec:limitations}: many flagged
disagreements originate in the mapping task itself, not the pipeline.

\begin{enumerate}
\itemsep1pt\parskip0pt\parsep0pt
    \item \textbf{Genuine DPV gaps.} \textit{Job Titles} $\to$
    \texttt{pd:Job} is the closest available class but notably
    broader than the Bearer type; no narrower class exists.
    \textit{Lastname} $\to$ \texttt{pd:Identifier} persisted after
    reformulated prompting, since DPV's \texttt{pd:Name} definition
    (``given name or nickname'') excludes surnames and no
    \texttt{pd:Surname} class exists.
    \item \textbf{Upstream scanner ambiguity.} \textit{Emails} $\to$
    email content could equally denote a contact list
    (\texttt{pd:EmailAddress}); \textit{Employee Files} $\to$
    \texttt{pd:Profile} is a composite HR container spanning financial
    and Article 9 health/union data that Bearer's own category label
    (``Professional Information'') had already miscategorized before
    the mapping LLM saw it.
    \item \textbf{Purpose-based risk blind spot.} \textit{Correct}
    verdicts on preference data (\textit{Accents},
    \textit{Acquaintances}, \textit{Browsing Behavior}) carry a caveat
    that \emph{inferential use} (e.g. profiling political opinion,
    per \emph{CJEU C-300/21 ``Austrian Post''}) could 
    trigger Article 9 in a way a category-based mapping cannot capture.
\end{enumerate}

These patterns show, independently of pipeline correctness, where
DPV-PD's categories and Bearer's taxonomy reach their representational
limits.

\subsection{Gate Summary and Summative Evaluation}
\label{sec:gate-summary}

Table~\ref{tab:formative} summarizes each gate, the corresponding verification procedure, and its outcome.

\begin{table}[H]
\centering
\small
    \vspace{-5pt}
\caption{Formative gate outcomes.}
    \vspace{-5pt}
\label{tab:formative}
\renewcommand{\arraystretch}{1.0}
\begin{tabular}{@{}l p{6.75cm}@{}}
\toprule
\textbf{Gate} & \raggedright\textbf{Observed outcome}\tabularnewline
\midrule
\textbf{G1} & \raggedright\textbf{Pass}. Three shapes evaluated 118 policies; zero violations.\tabularnewline
\textbf{G2} & \raggedright\textbf{Pass}. Zero Critical; 3 Important and 2 Minor findings were non-blocking.\tabularnewline
\textbf{G3} & \raggedright\textbf{Pass}. CQ-01--05 returned all five expected row counts.\tabularnewline
\textbf{G4} & \raggedright\textbf{Pass}. Nine annotators; raw agreement 0.721; AC2 0.877.\tabularnewline
\textbf{G5} & \raggedright\textbf{Pass}. OG 0.941; CR 0.780; 10/79 mappings flagged for revision.\tabularnewline
\bottomrule
\end{tabular}
    \vspace{-5pt}
\end{table}

\paragraph{Internal summative checks.} Contribution completeness, a
clean-environment rerun, and author self-assessment checked process
transparency, not independent effectiveness.

\subsection{Threats to Validity}

    \noindent
    \textit{Internal validity}: the LLM classification may
    contain systematic biases not caught by an automated judge scoring
    the same model's own output. Mitigation: a separate
    human-annotation round (Gate G4) plus review of flagged mappings.

    \vspace{5pt}
    \noindent
    \textit{External validity}: the mapping is limited to Bearer
    CLI data types (122 entries); other scanners use different
    taxonomies. Mitigation: designed for adaptation to other datatype
    taxonomies (Section~\ref{sec:review}), but unevaluated.

    \vspace{5pt}
    \noindent
    \textit{Construct validity}: DPV coverage is incomplete;
    some Bearer types (technical identifiers) have no DPV match.
    Mitigation: these cases are explicitly flagged as coverage gaps and deferred to Layer~2.

    \vspace{5pt}
    \noindent
    \textit{Conclusion validity}: raw agreement, $\alpha$, and
    Gwet's AC1/AC2 disagree sharply on the same data
    (Section~\ref{sec:results-annotation}), so which statistic is
    foregrounded could shift the reliability conclusion. Mitigation:
    reported side by side rather than collapsed into one number.

\section{Limitations}\label{sec:limitations}
The pipeline calls the hosted mapping LLM without a temperature or seed
override, so outputs can vary; self-hosting remains future work. This
matters because structured-output generation has a measurable reliability
ceiling~\cite{yang2026structeval}, and format constraints can degrade
reasoning relative to free-form generation~\cite{tam2024format}. Both are
relevant to the 8\% \texttt{flag\_revision} rate reported in
Section~\ref{sec:mapping-coverage}. Bearer CLI's Elastic License 2.0 does not affect our artifact, which
consumes only its output taxonomy. DPV coverage is also incomplete.
Esteves and Rodr\'iguez-Doncel~\cite{esteves2024} find 31 of 57 GDPR
informational items represented, and our Gate G4 annotation
(Section~\ref{sec:results-annotation}) identifies concrete \texttt{pd:*}
gaps as candidate DPV extension points. The motivating developer
survey~\cite{matos2025} is Brazil-specific, although non-Brazilian studies
corroborate it~\cite{senarath2018,hadar2018}.

The annotation study evaluates a fixed set of Bearer-to-DPV-PD mappings,
not downstream developer decisions. Its 43 \texttt{trivial\_pass} entries
(exact-label matches, Section~\ref{sec:mapping-coverage}) were reviewed
only by the pipeline's authors, not human-annotated; we considered this
low-risk given the narrow check (Bearer labels matched against DPV-PD),
not because it received independent verification. The Juice Shop
demonstration --- a clone-by-URL scan returning GDPR-provision badges with
file/line/field provenance --- shows the pipeline works as an artifact,
but whether it reduces privacy-review effort or errors relative to a
baseline remains untested, requiring a task-based developer study.

\section{Conclusion and Future Work}\label{sec:conclusion}
PrivDev Layer~1 bridges security-scanner datatype signals to DPV-PD
categories and machine-readable links to potentially relevant GDPR provisions. Retrieval-grounded LLM proposals are constrained by
deterministic encoding and SHACL and RAG-quality routing (RQ$_1$). The
SPARQL-queryable graph passed five formative gates, providing structural,
retrieval-quality, and human evidence of mapping plausibility, though
plausibility remains less stable for ambiguous or ontology-limited cases
(RQ$_2$).
Internal checks support process transparency but are not independent
validation. Ambiguous scanner labels, datatype-only evidence, and DPV-PD
coverage constrain precision (RQ$_3$); Layer~1 supports developer-facing
privacy review rather than establishing legal compliance. The FEDS-structured
convergent-evidence protocol itself --- formative structural gates, an
LLM-judge signal, and multi-statistic human annotation --- is reusable
beyond this mapping task for other LLM-assisted vocabulary-alignment work.
We view this as an ongoing maintenance concern rather than a one-time
check: CI-integrated re-mapping can catch drift as \emph{privacy debt}
accumulated over code evolution, before it compounds.
Future work will extend the bridge to Layer~2 (CWE) and Layer~3 (OWASP)
mappings via a planned PrivDev Gateway, spot-check \texttt{trivial\_pass}
mappings, grow the annotator panel, and integrate confidence-weighted
gating into CI.

\phantomsection
\section*{Artifact Availability}\label{sec:artifacts}
All artifacts are available at
\url{https://github.com/aisepucrio/privdev-layer1-artifact}
\cite{GitHubRepo}.


\section*{Acknowledgements}
This research was partially funded by the Brazilian funding agencies CAPES/COFECUB (Grant 88881.879016/2023-01 and Ma1036/24), CNPq (Grant 406089/2025-6), FAPESP (Grant 2023/00811-0), and FAPEMIG (Grant APQ-01488-24). 
We also acknowledge the Brazilian company Stone\footnote{\url{https://www.stone.com.br/}} for their financial support.

We thank the law professionals who participated in our study and shared their perspectives about PrivDev.

\clearpage
\balance
\bibliographystyle{ACM-Reference-Format}
\bibliography{ref}






\end{document}